\documentclass{article}

\usepackage{geometry}
\usepackage{graphicx} 
\usepackage{amsmath} 
\usepackage{amsfonts} 
    \DeclareMathSymbol{\shortminus}{\mathbin}{AMSa}{"39}
\usepackage{graphicx} 
\usepackage{authblk}  
\usepackage{setspace}
\usepackage{soul}
\usepackage{color} 
\usepackage{bbold}
\usepackage{multirow} 

\usepackage{algorithm}  
\usepackage{algpseudocode} 

\usepackage[utf8]{inputenc}
\usepackage[backend=biber,style=nature,sorting=none]{biblatex}

\title{Photonic Matrix Multiplication Circuit Based on Double Racetrack Resonator Building Blocks}
\author[1]{Hussein Talib}
\author[1]{Phillip D. Sewell}
\author[1]{Ana Vukovic}
\author[1]{Sendy Phang}
\affil[1]{George Green Institute for Electromagnetics Research, Faculty of Engineering, University of Nottingham, Nottingham, NG7 2RD, UK}

\affil[1]{\text{hussein.talib@nottingham.ac.uk, Sendy.Phang@nottingham.ac.uk}}

\date{July 2025}

\begin{document}
\setstretch{1.5}
\maketitle

\newpage
\begin{abstract}
This paper presents a novel design framework for photonic matrix multiplication based on programmable photonic integrated circuits using double racetrack (DRT) resonators as building blocks. Here, we analytically demonstrate that the transfer function of the DRT resonator building block resembles that conventional building blocks, such as directional couplers and MZI, making it suitable for building programmable circuits that handle complex matrix calculations. Using this new DRT resonators building block, a 3-by-3 photonic processor is implemented and validated through full-wave Finite Element Method (FEM) simulations, and scalability is further analysed using hybrid FEM-circuit modelling. Additionally, we implement a low-pass filter as a non-unitary system example, showcasing the flexibility of the approach. Results confirm high fidelity between simulated and analytical models, supporting the viability of  DRT resonators for reconfigurable photonic circuits. We believe that the proposed DRT resonator building blocks have the potential to complement and integrate with other previously reported blocks, thereby enhancing the fidelity and expanding the application scope of programmable photonic integrated circuits, particularly for all-optical signal processing in communication systems and for integration within microwave photonics platforms targeting emerging telecommunications technologies.
\end{abstract}
\newpage
\section{Introduction}
Photonic Integrated Circuits (PICs) are posed to become fundamental components in the different fields, including optical communications, sensing, and computing \cite{Zhuang2015,Hu2017,Poulton2019,zhang2020,Perez2017,Carolan2015,Najjar_Amiri2024}.
The increased diversity and complexity of photonic applications have led to the development of custom-designed PICs with advanced performance capabilities, which are a critical factor in deriving progress in photonic technologies. Classical design approaches, known as Application Specific Photonic Integrated Circuit (ASPIC), depend on prior knowledge and requirements to create devices with specific functions, offering limited capability for improvement and optimisation \cite{fandino2017monolithic}. More recent approaches, including inverse design approaches, offer significantly greater design flexibility in terms of functionality and performance by using specific computational and optimisation algorithms such as Genetic Algorithms (GA) 
\cite{Sewell2007,Vukovic2010,Piggott2015,phang2024,Najjar_Amiri2024}. 
\hfill \newline \hfill \newline
However, the optimisation, functionality, and overall performance of a photonic design are closely linked to the number of tunable parameters available. While a higher number of tunability enables the implementation of more complex and arbitrary operations, it also increases the complexity of the optimisation process, demanding greater computational resources and time \cite{Piggott2015,Lu2013,Zhang2022,Jia2018,Najjar_Amiri2024}.
The ultimate goal in PIC design is to realise a flexible, reprogrammable platform whose functionality is governed by its topology, thereby eliminating the need for a full redesign for each new application. In recent years, programmable PICs constructed from building blocks, such as beam splitters, Mach–Zehnder Interferometers (MZI), Ring resonators, Directional Coupler (DC), tuneable couplers, Double Mach–Zehnder interferometers (DMZI), have been suggested as a promising solution to address this challenge \cite{Miller:13,Perez2017,Capmany2020,Bogaerts2020,Yi2021,mosses2023design,Talib2025,multifunctional,dual-drive,valdez2025high,Xu2022,Perez2020, Najjar_Amiri2024}. The circuits are developed based on mathematical linear-system decomposition methods \cite{reck_1994}. Using such methods, a photonic circuit capable of performing any linear input–output unitary operation can be implemented in a structured topology, such as triangular \cite{reck_1994}\cite{Miller:13}, rectangular \cite{Clements2016,Fldzhyan:20,Marchesin:25}, diamond \cite{Diamond:20}, and Bokun \cite{Addressing:23}.  
\hfill \newline \hfill \newline
Each building block offers distinct advantages and/or trade-offs in spectral control, compactness, and tuning complexity. While DCs are simpler to design and effective, it often require a longer optical propagation distance to achieve the required phase change and coupling \cite{Talib2025}. To address this, a building block based on crossbar microring resonators was introduced \cite{Yi2021}. However, such a building block requires complex phase control schemes and suffers from limited free spectral range due to the physical size of the rings. Alternatively, Mach–Zehnder interferometer (MZI) is commonly used as a building block due to its tunability and ease of integration with other components \cite{Bogaerts2020,Perez2017}. However,  MZIs have limitations in terms of achievable extinction ratio and overall signal fidelity \cite{valdez2025high}. To overcome this, \cite{valdez2025high} proposed a Doubly Mach-Zehnder Interferometer (DMZI) where the high extinction ratios are improved. However, it increases system complexity and the number of tuning parameters \cite{valdez2025high}.  
\hfill \newline \hfill \newline
In this work, we propose a new building block based on Double Racetrack (DRTs) resonators. This building block offers a compact design, finer control of tuning parameters, and improved spectral control \cite{rabus2020integrated}. We demonstrate this DRT building block on the rectangular circuit topology \cite{Talib2025}. The circuit design framework using the new DRT resonators building blocks is validated on a 3-by-3 matrix multiplication demonstrator and then an all-optical low-pass filter. By employing such a structured circuit, it becomes possible to build a single circuit that can be reconfigured depending on the target task, enhancing flexibility and efficiency in different applications. Throughout this work, modelling is achieved using the full-3D Finite Element Method (FEM) from the commercial package COMSOL \cite{comsol2023}. Our findings show that the proposed DRT resonator building block enables a compact and tunable photonic circuit design with improved spectral control and is well-suited for narrow-band applications such as reconfigurable all-optical filter implementations.
\hfill \newline \hfill \newline
This paper is organised as follows: Section 2 details the transfer function of the DRT resonator building block used in the circuit design framework. Section 3 described the circuit design framework and the tuning parameters of the  DRT resonator building block. Section 4 demonstrates the capability of the proposed building block to implement a unitary matrix multiplication operation, analysis of its scalability , and a practical application of the analogue low-pass all-optical filter. Section 5 summarises the work.

\section{Double racetrack resonator building block}
\label{Analysis}
\begin{figure}[tbp]
    \centering
    \includegraphics[width=0.8\columnwidth]{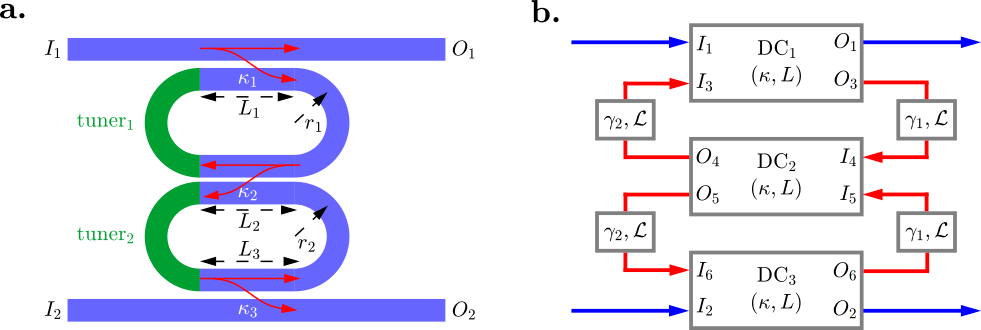 }
    \caption{(\textbf{a}) Physical layout of  DRT resonator building block. (\textbf{b}) Equivalent circuit model.}
    \label{fig: racetrack_resonator_diagram}
\end{figure}

In this section, we derive the transfer function of the proposed DRT resonator, which forms the fundamental building block of the circuit. Figure \ref{fig: racetrack_resonator_diagram}\text{(a)} illustrates the layout of the building block showing two RT resonators denoted by input ports $(I_{1,2})$, output ports $(O_{1,2})$, bent radius $(r_{1,2})$, interaction lengths $(L_{1,2,3})$, refractive index control elements $(\text{tuner}_{1,2})$, and coupling coefficient $(\kappa_{1,2,3})$. For identical RT resonators, the equivalent circuit model of this layout is shown in Fig. \ref{fig: racetrack_resonator_diagram}\text{(b)} where it consists of input ports $(I_{1,\dots,6})$, output ports $(O_{1,\dots,6})$, directional couplers $(\text{DC}_{1,2,3})$ characterised by coupling coefficients $\kappa$ and interaction length at coupler region $L$, arc length $\mathcal{L}$, propagation constants $\gamma$, and feedforward/feedback loops representing the resonators highlighted by red curved arrows. Assuming for low overall loss, directional coupler building blocks can be modelled as,
\begin{equation}
    \label{eq:oioj}
    \begin{bmatrix}
    O_i \\
    O_j
    \end{bmatrix}
    =
    \textbf{T}^{\text{DC}}
    \begin{bmatrix}
    I_i \\
    I_j
    \end{bmatrix}, \quad \text{for } (i,j) \in \{(1,3), (4,5), (6,2)\},
\end{equation}
where
\begin{equation}
    \label{T_dc}
    \textbf{T}^{\text{DC}}
    =e^{-\text{j}\gamma_1 L}
    \begin{bmatrix}
        \cos(\kappa L) & \text{j}\sin(\kappa L)\\
        \text{j}\sin(\kappa L) & \cos(\kappa L)
    \end{bmatrix}.
\end{equation}
The phase delay of the optical carrier resulting from propagation on the bend waveguide is modelled by,  
\begin{equation}
\label{eq:I4I5}
 \begin{bmatrix}
     I_4\\I_5
 \end{bmatrix}
 = e^{-\gamma_1 \mathcal{L}}
 \begin{bmatrix}
     O_3\\O_6
 \end{bmatrix} \text{ and }
\begin{bmatrix}
    I_3\\I_6
\end{bmatrix}
= e^{-\gamma_2 \mathcal{L}}
\begin{bmatrix}
    O_4\\O_6
\end{bmatrix}.
\end{equation}
where the propagation constants $\gamma_1=\text{j}\beta$, $\gamma_2=\text{j}(\beta+\Delta\beta)$, physical length of the bent waveguide $\mathcal{L}=\pi r$, $\beta$ is the propagation constant, and $\Delta\beta$ is the change of propagation constant due to the small change in the refractive index in $\text{tuner}_1$ and $\text{tuner}_2$.
\newline
 
Simultaneously solving eqs. $(\ref{eq:oioj})$ to $(\ref{eq:I4I5})$, the transfer functions of the DRT resonator building block are,
\begin{align}
\label{O1_I1}
\begin{split}
    \frac{O_1}{I_1} = \frac{O_2}{I_2} &= \vartheta \frac
    {(CS^2-C^3)(1 - \xi_1 \xi_2)^2 + C S^2 \xi_1 \xi_2}
    {S^2 - C^2 (1 - \xi_1 \xi_2)^2 }, \\
    \frac{O_1}{I_2} = \frac{O_2}{I_1} &= \vartheta \frac
    {S^3\xi_1 \xi_2}
    {S^2 - C^2 (1 - \xi_1 \xi_2)^2 }.
\end{split}
\end{align}
The assumption of negligible loss is valid for waveguides with sufficiently large bend radii, where bending-induced losses are minimal, as well as high fabrication fidelity (i.e. low sidewall roughness) and low material absorption, both of which are achievable with current silicon photonic fabrication technology. In eq. (\ref{O1_I1}), $\vartheta = e^{-\text{j} \beta L}$, $C = \cos{(\kappa L)}$, $S = \mathrm{j}\sin{(\kappa L)}$, $\xi_1 = e^{-\text{j}\beta\mathcal{L}}$, and $\xi_2~=~e^{-\text{j}\Delta\beta\mathcal{L}}$. Under near resonance operation, the term $\xi_1$ approaches unity. For the basic DRT structure with the tuners switched off, $\Delta \beta = 0$, thus $\xi_2=1$, leading to,  
\begin{align}
    \label{eq:O1_I1_s}
    \begin{split}
        \frac{O_1}{I_1} = \frac{O_2}{I_2} = \vartheta \cos(\kappa L),\\
        \frac{O_2}{I_1} = \frac{O_1}{I_2} = \text{j}\vartheta \sin(\kappa L).
    \end{split}
\end{align}
or in the matrix form,
\begin{equation}
\label{eq:T_DRT}
\textbf{T}^{\text{DRT}} = \vartheta
\begin{bmatrix}
    \cos \theta  & \text{j}\sin \theta \\ 
    \text{j}\sin \theta & \cos \theta
\end{bmatrix}, \quad \text{with } \theta = \kappa L.
\end{equation}
It can be seen that, under near-resonance operation and assuming minimal loss, the DRT building block exhibits a transfer function, eq. (\ref{eq:T_DRT}), that closely resembles that of a directional coupler.

\section{Circuit design framework}
\label{framework}
\begin{figure}[tbp]
    \centering
    \includegraphics[width=0.8\columnwidth]{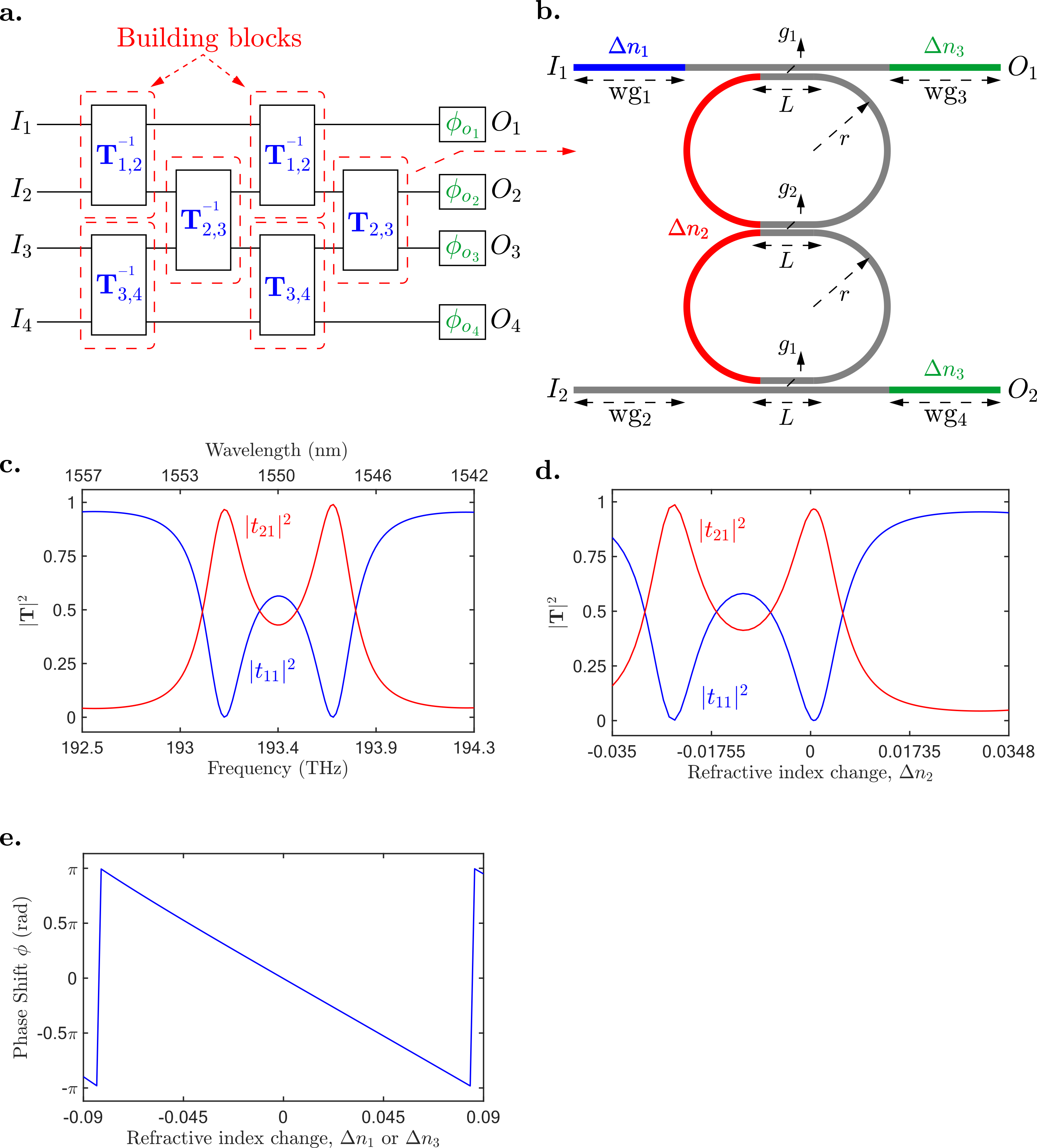}
    \caption{(\textbf{a}) An example of 4-by-4 PIC with required building blocks and output phase shifts $\phi_o$. (\textbf{b}) DRT geometry layout and its design parameters. (\textbf{c}) Spectrum of the magnitude of the bar transmission $|t_{11}|^2$ and cross transmission $|t_{21}|^2$, for $\Delta n_{1,2,3}=0$.
    (\textbf{d}) Spectrum $|t_{11}|^2$ and $|t_{21}|^2$ as a function of $\Delta n_2$ of the arcs of DRT. (\textbf{e}) Phase shift $\phi$ as a function of $\Delta n_1$ or $\Delta n_3$.}
    \label{fig: racetrack_resonator_methodology_v3}
\end{figure}

In this section, we present a design framework for a PIC based on a DRT resonator. The process begins by considering a target unitary matrix $\mathbf{U}$, which is to be synthesised into a physical PIC. In cases where a non-unitary matrix is given, unitary matrix recovery techniques, e.g. Singular Value Decomposition (SVD) \cite{Miller:13}, Cosine Sine Decomposition (CSD) \cite{Talib2025}, or  Genetic Algorithm (GA) \cite{talib2025owtnm} can be used prior to the design framework described here. For specific details of the CSD approach, readers are referred to \cite{Talib2025}.
\hfill \newline \hfill \newline
The design framework comprises two stages: circuit synthesis and building block design using a photonic simulator. Circuit synthesis is achieved by applying a nulling process to the matrix $\mathbf{U}$ to determine the design parameters required for each building block, using the transfer function of the chosen building block, i.e. DRT resonators, and the phase shift at the output of the circuit $\phi_o$ as illustrated in the 4-by-4 circuit example in Fig. $\ref{fig: racetrack_resonator_methodology_v3}$(a). The nulling process depends on the circuit topology of the circuit; for the rectangular topology used in the present work, we employ the decomposition algorithm developed by Clements, \textit{et al}; for further details of this process, readers are referred to \cite{Clements2016}. To enable control of the relative phase of the incoming wave between the two input ports, the DRT building block design shown Fig. \ref{fig: racetrack_resonator_diagram}(a) is modified by incorporating a tunable phase delay at the upper input port of the DRT resonators building blocks, implemented as a short waveguide section with a tunable refractive index, see Fig. \ref{fig: racetrack_resonator_methodology_v3}(b). Taking this pre-phasing tuning into account, the updated transfer function of the DRT building block is given by,  
\begin{equation}
\label{eq:T}
\begin{bmatrix}$
$O_1 \\$
$O_2$
$\end{bmatrix}
=
\mathbf{T^{\text{DRT}}(\theta,\phi)}
\begin{bmatrix}$
$I_1 \\$ 
$I_2$
$\end{bmatrix},
\quad
\text{with }
\mathbf{T^{\text{DRT}}(\theta,\phi)} =
\begin{bmatrix}
e^{\text{j}\phi}\cos\theta & \mathrm{j}\sin\theta \\ 
\text{j}e^{\text{j}\phi}\sin\theta & \cos\theta
\end{bmatrix} \text{ and } \theta =  \kappa L.
\end{equation}

From eq. (\ref{eq:T}), it can be seen that the modified building block transfer function is characterised by two design parameters $(\theta,\phi)$, where $\theta=\kappa L$ corresponds to the effective phase of the DRT resonator, as described in Section \ref{Analysis}, and $\phi$ represents the phase-shift induced by the change in refractive index of the pre-phasing tuning section, i.e. $\phi(\Delta n_1)$. Furthermore, note that the common phase $\vartheta$ in eq. (\ref{eq:T_DRT}) has been removed, as it was applied equally to the output waves at both ports and its effect is compensated by the newly added tunable phase delay $\Delta n_3$, see Fig. \ref{fig: racetrack_resonator_methodology_v3}(b). Throughout this paper, we will omit the explicit dependence of $\textbf{T}^{\text{DRT}}(\theta,\phi)$ on $\theta$ and $\phi$ for notational simplicity.
\hfill \newline \hfill \newline
The circuit synthesis by nulling process decomposes the matrix $\textbf{U}$ into a sequence of building blocks $\textbf{T}^{\text{DRT}}_{p,q}$ and $\left(\textbf{T}_{p,q}^{\text{DRT}}\right)^{-1}$ and a diagonal matrix $\textbf{D}$ as \cite{Clements2016,Capmany2020},
\begin{equation}
\label{nulling_1}
\textbf{U} = 
\left(
\prod_{\substack{(p,q) \in S_L \\}} \left(\textbf{T}^{\text{DRT}}_{p,q}\right)^{-1}
\right)
\textbf{D}
\left(
\prod_{\substack{(p,q) \in S_R \\ }} \textbf{T}_{p,q}^{\text{DRT}}
\right).
\end{equation}

In eq. (\ref{nulling_1}), $\textbf{D} = \mathrm{diag}\left(e^{\text{j}\phi_{01}}, e^{\text{j}\phi_{02}}, \dots, e^{\text{j}\phi_{0N}}\right)$ is a complex-valued unit diagonal matrix representing the phase shifts at the output of the circuit; see Fig. $\ref{fig: racetrack_resonator_methodology_v3}$(a).The algorithm of the decomposition method by nulling is detailed in the Supplementary Material. The matrices $\textbf{T}^{\text{DRT}}_{p,q}$ and $\left(\textbf{T}^{\text{DRT}}_{p,q}\right)^{-1}$ denote the partitioned building blocks' transfer matrix, with $S_L$, $S_R$ representing the index ordering for the  matrices used to null the $(p,q)$-th row and column (with $p<q$) of the matrix $\textbf{U}$, as: 

\begin{align}
\label{eq:TMN}
\nonumber & \qquad 1 \quad \text{ } 2 \quad \text{ } \cdots \qquad \text{ } p \qquad \text{ } \cdots \quad \text{ }  q  \qquad \text{ }  \cdots  \quad \text{ } \quad \text{ } N \\
\mathbf{T}^{\text{DRT}}_{p,q} & = 
\begin{bmatrix}
1 & 0 & & \cdots & & \cdots & & \cdots & 0 \\
0 & 1 & & & & & & & \vdots \\
& & \ddots & & & & & & \\
\vdots & & & e^{\text{j}\phi} \cos\theta & & \text{j}\sin\theta & & & \vdots \\
& & & & \ddots & & & & \\
\vdots & & & \text{j} e^{\text{j}\phi} \sin\theta & & \cos\theta & & & \vdots \\
& & & & & & \ddots & & \\
0 & \cdots & & \cdots & & \cdots & & 0 & 1 \\
\end{bmatrix}
\begin{array}{c}
1 \\
2 \\
\vdots \\
p \\
\vdots \\
q \\
\vdots \\
N
\end{array}
\end{align}


\hfill \newline

The set of parameters ($\phi,\ \theta,\ \text{and},\ \phi_o$), obtained from the decomposition in eq. (\ref{nulling_1}) is now implementable as physical DRT building blocks, designed with a full-vectorial photonic simulator. For a realistic and realisable implementation, the PIC is developed on a standard silicon-on-insulator (SOI) platform, with a silicon layer thickness of 0.22 $\mu$m. The circuit is based on strip waveguides with a width of 0.4~$\mu$m, operating in a single mode at a wavelength of $\lambda_\text{op}= 1.548\text{ }\mu$m, and buried in silicon oxide (silica) material. The refractive index of silicon is $n_\mathrm{silicon} = 3.455$, and the refractive index of the silica is $n_\mathrm{silica} = 1.445$.
\hfill \newline

Full-vectorial 3D FEM \cite{comsol2023} is used to design and simulate each of the DRT building blocks. For design specificity, the straight waveguide sections at the input and output ports ($\mathrm{wg}_{1,2,3,4}$) each have a length of 8.42 $\mu$m. The gap between the racetrack resonator and the adjacent waveguide is $g_1=0.18\text{ }\mu$m, while the inter-racetrack resonator gap is $g_2= 0.06\text{ }\mu$m. To ensure accurate simulations while maintaining computational efficiency, an adaptive mesh refinement strategy \cite{comsol2023} was employed. Specifically, the high refractive index waveguide core region was meshed using a maximum element size of $\Delta = 64.015$ nm, while a coarser mesh with maximum element size of $\Delta = 153.06$ nm was used in the silica region.
\hfill \newline

As the transfer matrix model assumes operation near the resonance, the initial step in the design process is to ensure that the basic DRT structure resonates at the target operating frequency. FEM simulations were performed to optimise the basic design parameters under the condition $\Delta n_2=0$, from which it was determined that resonance occurs when identical racetrack resonators have a bend radius of $r=5.59\text{ }\mu$m and a straight waveguide sections of $L=4 \text{ }\mu$m. Figure \ref{fig: racetrack_resonator_methodology_v3}(c) shows the transmission spectrum of such DRT building block with $\Delta n_{1,2,3} = 0$. The spectrum exhibits two resonant peaks corresponding to the even $f_\text{even}=193.2$ THz and odd $f_\text{odd}=193.7$ THz modes, arising from mode splitting due to strong coupling between the two resonators. The fundamental resonant frequency of a single racetrack resonator is given by $f_0=\frac{1}{2}(f_\text{even}+f_\text{odd})=193.4$ THz. 
\hfill \newline

To realise the computed parameters ($\phi,\ \theta,\ \text{and},\ \phi_o$), the DRT building block is tuned via small changes in the refractive indices $(\Delta n_1,\Delta n_2,\ \text{and}\ \Delta n_3)$. Specifically, adjusting $\Delta n_2$ shifts the resonance frequency of the DRT, thereby tuning the transmission amplitude and enabling the desired value of $\theta$ to be achieved. Figure \ref{fig: racetrack_resonator_methodology_v3}(d) confirms this tunability, showing that the transmission at $\lambda_\text{op}= 1.5482\text{ }\mu$m can be varied from nearly unity to near zero. Adjusting $\Delta n_1$ allows for tuning of the transmission phase for input from port 1, enabling the desired phase shift $\phi$ to be achieved. Additionally, tuning $\Delta n_3$ corrects for any additional phase delay introduced by operating slightly off-resonance, thereby preserving phase coherence across the output channels while maintaining the desired transmission amplitude. Figure \ref{fig: racetrack_resonator_methodology_v3}(e) confirms that full phase tuning range from $-\pi$ to $\pi$ can be achieved through a small refractive index adjustment of $\Delta n_{1,3}<0.2$. Note that this refractive index change can be further reduced by increasing the length of the waveguide sections ($\mathrm{wg}_1$, $\mathrm{wg}_2$, and $\mathrm{wg}_3$). In the special case of building blocks located near the output (i.e., at the far right of the circuit), $\Delta n_3$ is also used to achieve the target output phase of $\phi_o$.

\section{Results and discussion}
This section presents the validation and application of the design framework proposed in Section \ref{framework} for the PIC using the DRT resonator as a building block. 

\subsection{Validation}
\label{Validation}
\begin{figure}
    \centering
    \includegraphics[width=0.8\columnwidth, height=0.9\textheight, keepaspectratio]{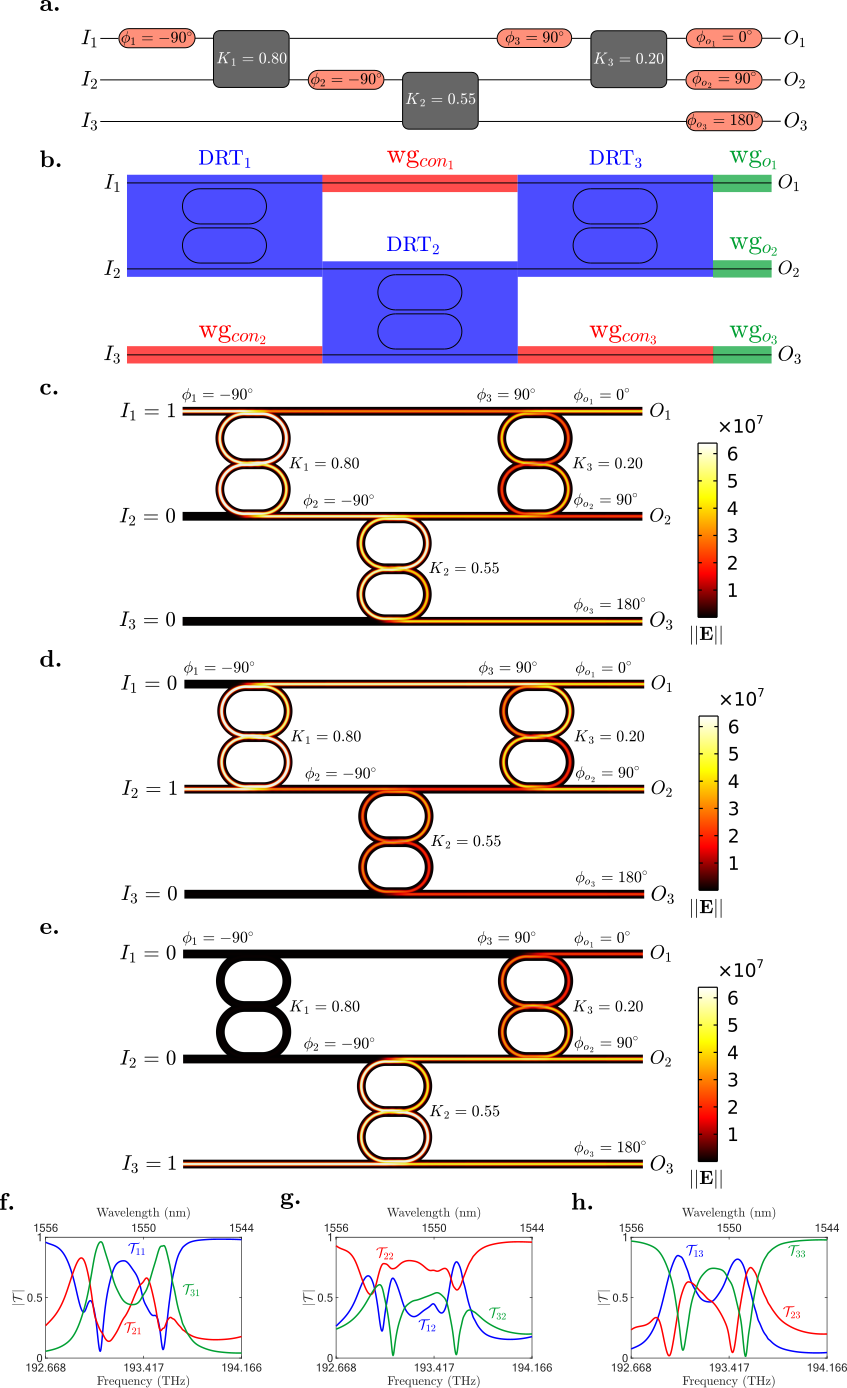}
    
    \caption{(\textbf{a}) Schematic of the practical realisation of the 3-by-3 unitary matrix. Coupling parameter $K$ and external
phase shift $\phi$ for each building block are calculated as in Section $\ref{framework}$. (\textbf{b}) Schematic of the 3-by-3 photonic
circuit where phase shift $\phi$ are realised by tuning $\Delta n_1$ and the coupling constant $K$
is realised by tuning $\Delta n_2$. (\textbf{c–e}) Full-wave 3D simulation of the designed 3-by-3 photonic circuit when (\textbf{c}) $I_1$ is
on, (\textbf{d}) $I_2$ is on, and (\textbf{e}) $I_3$ is on. (\textbf{f-h}) The transmission parameters $\mathcal{T}_{i,j}$ are entries of $\mathcal{T}$ as a function of frequency (wavelength), where $i,j$ are the output and input ports.}
    \label{results_discussion_v1}
\end{figure}

As an exemplary case to validate the circuit synthesis framework, consider a 3-by-3 unitary matrix $\textbf{U}$,
\begin{equation}
\label{U}
\mathbf{U} =
\begin{bmatrix}
0.6666 & \shortminus0.6666 & \shortminus0.3350 \\
0.3333 & 0.6666 & \shortminus0.6670 \\
\shortminus0.6691 & \shortminus0.3299 & \shortminus0.6662 \\
\end{bmatrix} .
\end{equation}
\newline
\hfill \newline
Following the nulling process in Section \ref{framework}, the design parameters ($\phi,\ \theta,\ \text{and},\ \phi_o$) for each of the building blocks making up for the circuit in eq. (\ref{U}) are obtained. The parameters of the phase angles $\phi$, the output phases $\phi_{o}$ and the coupling parameters $K = \sin^2\theta$ for each building blocks are shown in Fig. \ref{results_discussion_v1}(a). Then, FEM simulations are performed to determine the required refractive index tuning $\Delta n_i$ needed to achieved the specified design parameters ($\phi,\ K,\ \text{and},\ \phi_o$). Additionally, we  adjust the refractive index of each interconnecting waveguide $\text{wg}_{\text{con}_i}$ and output waveguide $\text{wg}_{\text{o}_i}$ to correct for the signal's propagation phase, see Fig. \ref{results_discussion_v1}(b), thereby maintaining coherence across the building blocks. It is important to note that the decomposition approach assumes a direct cascading of building blocks, whereas physical realisation requires interconnecting waveguide between them. Table \ref{table_drt} summarises all these parameters.
\hfill \newline
\begin{table}[tb]
\centering
\caption{Summary of the circuit parameters for the demonstrator case (\ref{U}).}
\begin{tabular}{|c|c|c|c|c|c|c|c|c|c|c}
\hline
Components & $\phi_i$ & $K$ & Phase adjustment & $\phi_{o_{i}}$ & $\Delta n_1$ & $\Delta n_2$ & $\Delta n_3$ & $\Delta n_{\text{wg}_{con_i}} $ & $\Delta n_{\text{wg}_{out_i}} $\\
\hline
{DRT$_1$} & -$90^\circ$ & $0.80$ & -$99.8^\circ$ & - & 0.0427 & 0.0032 & 0.0474 & - & -\\
\hline
{DRT$_2$} & -$90^\circ$ & $0.55$ & -$83.1^\circ$ & - & 0.0427 & 0.0052 & 0.0394 & - & -\\
\hline
{DRT$_3$} & $90^\circ$ & $0.20$ & -$57.6^\circ$ & - & -0.0428 & 0.0102 & 0.0272 & - & -\\
\hline
$\text{wg}_{\text{con}_1}$ & - & - & -$11.2^\circ$ & - & - & - & - & 0.0050 & -\\
\hline
$\text{wg}_{\text{con}_2}$ & - & - & -$11.2^\circ$ & - & - & - & - & 0.0050 & -\\
\hline
$\text{wg}_{\text{con}_3}$ & - & - & -$11.2^\circ$ & - & - & - & - & 0.0050 & -\\
\hline
$\text{wg}_{\text{out}_1}$ & - & - & - & $0^\circ$ & - & - & - & - & 0\\
\hline
$\text{wg}_{\text{out}_2}$ & - & - & - & $90^\circ$ & - & - & - & - & -0.0428\\
\hline
$\text{wg}_{\text{out}_3}$ & - & - & - & $180^\circ$ & - & - & - & - & 0.0858\\
\hline
\end{tabular}
\label{table_drt}
\end{table}


Using these realisable parameters, the entire circuit is simulated using FEM. Figures \ref{results_discussion_v1}\text{(c-e)} show the optical signal intensity as it propagates through the circuit, when each input port is individually excited: (c) with $I_1$ activated, (d) with $I_2$ activated, and (e) with $I_3$ activated. From these simulations, the transmission parameters, at the operating wavelength, $\lambda=1.5482\ \mu\text{m}$, are obtained as
\begin{equation}
\label{U_FEM}
\textbf{O}=\mathcal{T}\textbf{I}, \quad \text{with }
\mathbf{\mathcal{T}_\text{FEM}} =
\begin{bmatrix}
0.6547\angle 3.92^\circ & 0.6513\angle 177.39^\circ & 0.3232\angle \shortminus178.08^\circ \\
0.3319\angle 13.56^\circ & 0.6544\angle\shortminus1.42^\circ & 0.6485\angle 177.26^\circ \\
0.6392\angle \shortminus170.39^\circ & 0.3249\angle\shortminus174.32^\circ & 0.6693\angle 177.65^\circ
\end{bmatrix}.
\end{equation}
We observe some discrepancies between the matrix elements of the target eq. (\ref{U}) and the FEM-simulated transfer matrix in eq. (\ref{U_FEM}). These are primarily due to intrinsic losses in the resonators, leading to the reduction in the magnitude of $\mathcal{T}_\text{FEM}$ compared to the ideal unitary. Additionally, phase deviations are introduced by shifting in the propagation constant caused by waveguide bending and by inter-waveguides coupling. Moreover, refractive index tuning also perturbs the propagation constants; an effect which was not considered in the decomposition procedure. Collectively, these physical real-world effects contribute to the observed mismatch between the simulated and target matrices. 
\hfill \newline

To quantify this discrepancy, we compute the Normalised Square Error (NSE), which yields a value of 0.0074. The NSE is calculated as  $\text{NSE} =||\mathcal{T}-\textbf{U}||^2 /N$ where $||.||^2$ is the L2 norm and $N$ is the dimension of the matrix. While this error is low, it reflects the impact of practical constraints in a circuit based on DRT building blocks. This result suggests that further optimisation in DRT block design may be necessary to achieve more accurate circuit implementations. For completeness, Fig. \ref{results_discussion_v1}(f-h) shows the transmission spectra over a narrow frequency band from 193 THz to 194 THz. The spectra exhibit complex, multi-resonance profiles, which is expected given that the circuit consists of several resonant DRT blocks. Each block contributes a distinct resonance peak, with slight shifts in position arising from variations in refractive index tuning across the circuit. Overall, the circuit exhibits a 0.8 $\%$ bandwidth centered around the designated operational frequency of 193.7 THz (wavelength of 1.5482 $\mu$m). For comparison, an alternative implementation of the same circuit rectangular topology using DC building blocks was found to exhibit a bandwidth of 4$\%$ \cite{Talib2025}. As expected, the DRT-based circuit demonstrates a narrower operational bandwidth due to its resonant nature.
\begin{figure}[tbp]
    \centering
    \includegraphics[width=0.6\columnwidth]{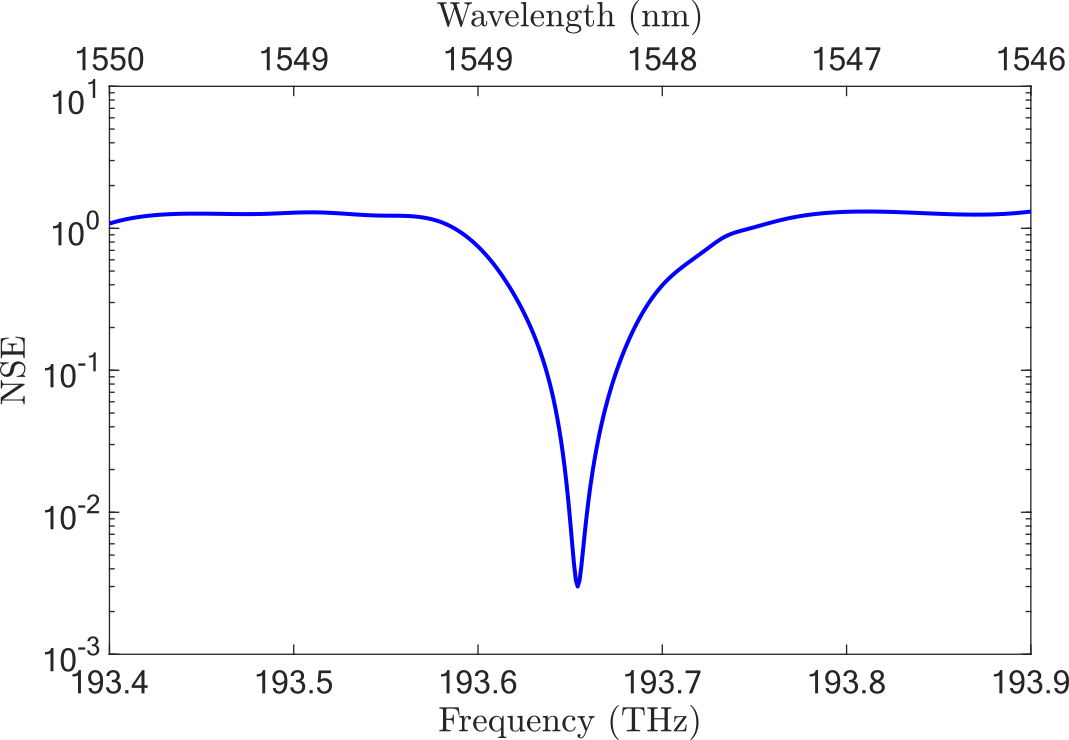}
    \caption{Normalised Square Error (NSE) as a function of frequency for the example 3-by-3 photonic integrated circuit.}
    \label{freq_vs_nse_3by3}
\end{figure}
\hfill \newline

Figure \ref{freq_vs_nse_3by3} shows the NSE as a function of frequency for the 3-by-3 configuration. A clear relationship is observed between the circuit implementation accuracy (NSE) and the the resonance nature of the circuit. The optimal accuracy occurs at the target operation frequency of 193.683 THz, where the NSE reaches its minimum. The half bandwidth, defined as the frequency range over which the NSE remains within half of its minimum value, is calculated to be 0.1050 THz, indicating a relatively narrow and selective frequency response. This narrow bandwidth highlights the system’s resolution capability and spectral precision, which are critical for applications such as optical filtering or sensing.

\subsection{Analysis of scalability}
\label{Analysis_of_scalability}
Having validated the synthesis framework, this section explores the scalability of the DRT building blocks for implementing larger circuits. Due to limitations in computational resources, full-wave 3D FEM simulations of the complete circuit were performed only for a 3-by-3 configuration to validate the design procedure. To model larger circuits, we employ a hybrid approach that combines full-wave FEM simulations of individual building blocks with circuit-based modeling to assemble the complete system. Specifically, each DRT block is individually optimised by tuning the refractive indices to match the parameters obtained from the nulling process, as previously described. The corresponding S-parameters generated by the FEM simulations are then imported into a circuit modeling tool \cite{Keysight2017} to construct the full photonic circuit. Since each building block is designed and simulated independently, this process can be parallelised to accommodate limited computational memory, enabling large-scale implementation.
\hfill \newline

The accuracy of this approach is evaluated by comparing the transfer parameters, $\mathcal{T}$, of the complete circuit obtained from FEM  simulations with those using the hybrid FEM-circuit approach. For the same target matrix \textbf{U} in Section \ref{Validation}, the $\mathcal{T}$ parameter computed from the hybrid approach is,

\begin{equation}
\label{U_FEM-circuit}
\mathbf{\mathcal{T}}_{\text{hybrid}} =
\begin{bmatrix}
0.6559\angle 6.44^\circ & 0.6510\angle \shortminus179.02^\circ & 0.3232\angle \shortminus175.12^\circ \\
0.3303\angle 18.88^\circ & 0.6555\angle 5.98^\circ & 0.6489\angle \text{-}174.88^\circ \\
0.6396\angle \shortminus169.52^\circ & 0.3251\angle \shortminus173.45^\circ & 0.6697\angle \shortminus179.97^\circ
\end{bmatrix}.
\end{equation}
By comparing eqs. (\ref{U_FEM}) and (\ref{U_FEM-circuit}), we observe an agreement between the two approaches, confirming the accuracy of the proposed hybrid approach. The NSE between $\mathbf{\mathcal{T}}_{\text{hybrid}}$ and the target eq. (\ref{U})  is calculated to be 0.0132. 
\hfill \newline
\begin{figure}[tbp]
    \centering
    \includegraphics[width=0.9\columnwidth]{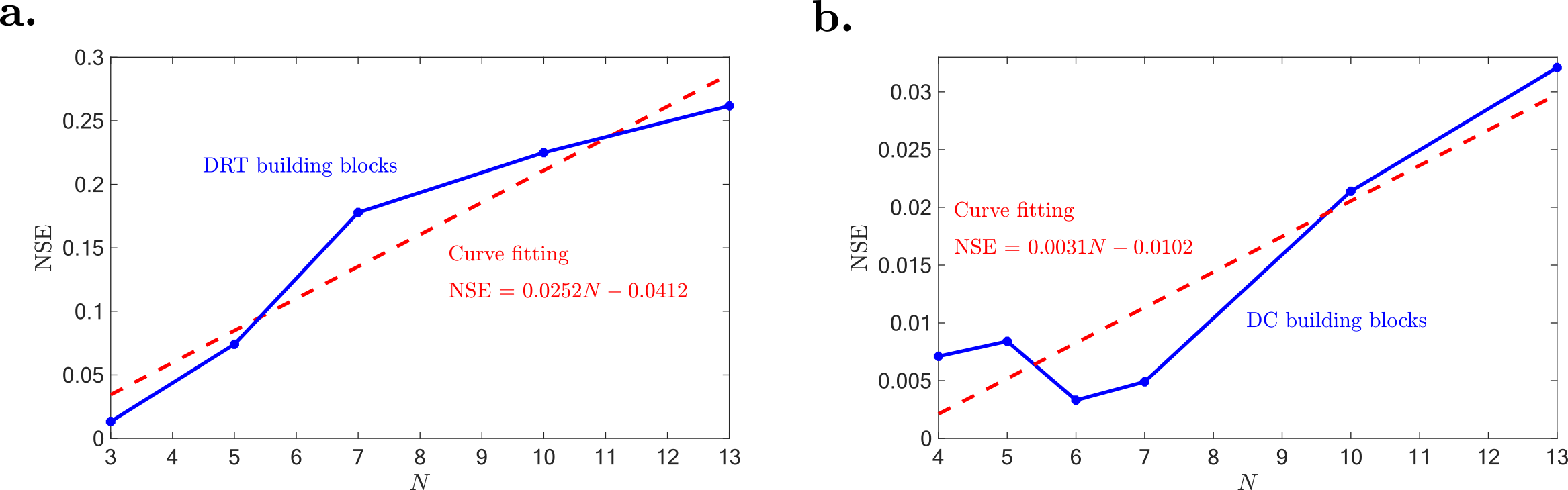}
    \caption{(\textbf{a}) NSE vs different sizes of PIC using DRT building blocks. (\textbf{b}) NSE vs different sizes of PIC using DC building blocks  \cite{Talib2025}.}
    \label{scalability_drt_vs_dc}
\end{figure}

The hybrid approach is now used to model circuits corresponding to different sizes of the target matrix $\textbf{U}$, including $N = 3,5,7,10,$ $\text{and}$ $13$. Figure \ref{scalability_drt_vs_dc}$(\text{a})$ shows that the NSEs of $\mathcal{T}$ increases with the size $N$ of the unitary matrix $\textbf{U}$. Linear interpolation of these NSE values shows an increase of approximately 0.0252 NSE units per increment in $N$ as shown in Fig. \ref{scalability_drt_vs_dc}\text{(a)}. This behaviour is expected due to the accumulation of slight inaccuracies between the decomposition and physically realisable circuit implementations, as described in the previous Section \ref{Validation}. For comparison, circuits based on DC building blocks \cite{Talib2025} exhibit an increase in NSE of approximately 0.0031 unit per $N$ as shown in Fig. \ref{scalability_drt_vs_dc}\text{(b)}. Consequently, this shows that the PIC based on DC is more scalable than the DRT-based design. 
\hfill \newline

\subsection{Analogue optical signal filter implementation}
\label{Analogue optical signal filter implementation}
To demonstrate the broader applicability of the proposed design framework, we consider a general system that is originally non-unitary. Since the framework is designed to implement unitary target systems, we apply a unitary recovery technique based on a Genetic Algorithm (GA) \cite{talib2025owtnm}, thereby recasting the original non-unitary problem into a unitary one. This transformation enables the application of the nulling-based photonic circuit design procedure described in the present work.
\hfill \newline

For this demonstration, we consider a low-pass filter system, whose transfer function, defined in the Laplace domain $s$, as \cite{Anufriev:22},
\begin{equation}
\label{laplace}
H(s) = \frac{y(s)}{x(s)} = \frac{1}{1 + \Gamma s},
\end{equation}
where $x$ and $y$ represent the input and output signals, respectively, and $\Gamma={1}/(2\pi f_H)$ is a filter parameter determined by target frequency cut-off $f_H$ of the low-pass filter. It is important to note that generally, any linear signal processing system with a known transfer function can be implemented using the design methodology described in this work.
\hfill \newline

To enable implementation, the transfer function eq. (\ref{laplace}) is reformulated into a standard IIR filter format in the time domain by applying the bilinear Z-transform, $s \leftarrow \frac{2}{\Delta T} \left( \frac{1 - z^{-1}}{1 + z^{-1}} \right) $, as
\begin{equation}
\label{time_domain}
y(n) = W_1 x(n) +   W_1 W_2 y(n-1) + W_1 x(n-1) \\
\end{equation}     
where 
\begin{equation}
\label{parameters_filter}
W_1 = \left( \frac{2\Gamma}{\Delta T}+1 \right)^{-1}; \quad W_2 = \frac{2\Gamma}{\Delta T}-1.
\end{equation}
The IIR filter in eq.  (\ref{time_domain}) shows that the current output signal $y(n)$ is recursive,  as it is a linear combination of the current input $x(n)$, a delayed input $x(n-1)$, and a feedback term from the previous output $y(n-1)$. The delay is governed by the time delay parameter $\Delta T$, with $\Delta T	\ll \Gamma$.  
\begin{figure}[tbp]
    \centering
    \includegraphics[width=0.7\columnwidth]{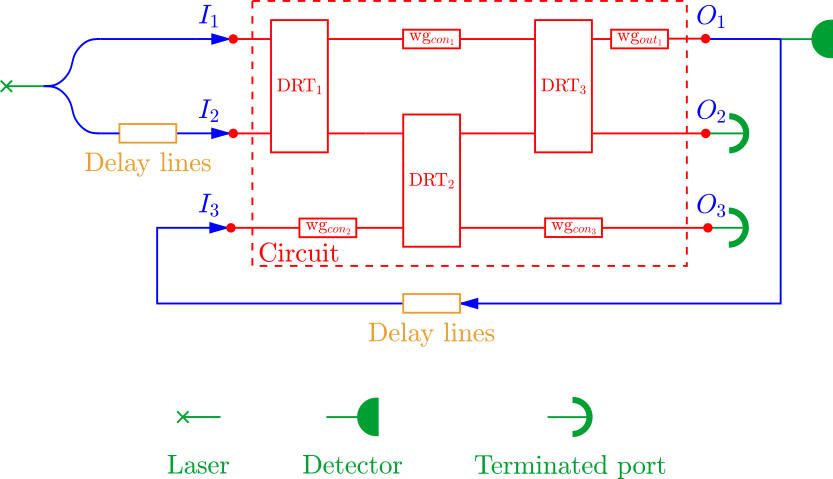}
    \caption{Illustration of the implementation of an all-optical filter based on a programmable PIC using DRT building blocks.}
    \label{iir_filter_diagram_circuit_v1}
\end{figure}
\hfill \newline

An analogue optical signal processor implementing the weights $W_1$ and $W_2$ is designed and modelled using the proposed framework. Figure~\ref{iir_filter_diagram_circuit_v1} illustrates the photonic circuit implementing this IIR filter. The input and output optical signals are split and delayed, practically achieved using physical optical delay lines or refractive index tuning, to produce the required time-shifted signals $x(n-1)$ and $y(n-1)$. The IIR filter circuit is shown in Fig. \ref{iir_filter_diagram_circuit_v1} and the part of the circuit highlighted in red implements the filter weights, and consists of three DRTs interconnected by waveguides, $\{\text{wg}_{\text{con}_1}, \text{wg}_{\text{con}_2}$, and $\text{wg}_{\text{con}_3}\}$, leading to the output port through $\text{wg}_{\text{out}_1}$. 
\hfill \newline \hfill \newline
Specifically, for a filter design parameter of $\Delta Tf_H=0.04$, the IIR digital filter in eq. (\ref{time_domain}) in the matrix form is given by
\begin{equation}
    \label{matrix_filter}
    \textbf{Y} = \textbf{W}\textbf{X} \equiv
    y(n) =
    \begin{bmatrix}
        0.1116 & 0.1116 & 0.7767\\
    \end{bmatrix}
    \begin{bmatrix}
        x(n)\\
        x(n-1)\\
        y(n-1)
    \end{bmatrix}, 
\end{equation}
with $\textbf{W}$ represents the filter coefficients to be synthesised into the red-highlighted circuit in Fig.~\ref{iir_filter_diagram_circuit_v1}. As $\textbf{W}$ is a non-square, non-unitary matrix, it must be converted into a unitary matrix using the GA unitary recovery approach \cite{talib2025owtnm} to enable circuit synthesis by nulling. The resulting 3-by-3 unitary matrix $\textbf{U}$, in which the first row represents the filter coefficients $\textbf{W}$ padded by additional entries to satisfy the unitarity condition as, 
\begin{equation}
\label{M_iir}
\textbf{U}_{\text{iir}} = 
1.26
\begin{bmatrix}
 0.1116 & 0.1116 & 0.7767 \\
 \shortminus0.7146 & 0.3412 & 0.0537 \\
 0.3268 & 0.7078 & \shortminus0.1487 \\
\end{bmatrix}
\end{equation}
\hfill \newline

Following the same decomposition procedure described in Section \ref{framework}, we implemented this matrix as a programmable photonic circuit based on tunable  DRT resonator building blocks. The detailed design parameters are provided in Table \ref{table_iir_filter} which yields to the FEM simulated transmission matrix of $\mathcal{T}_\text{FEM}$,
\begin{equation}
\label{iir}
\mathcal{T}_\text{simulated}=1.26
\begin{bmatrix}
0.1486\angle 9.94^\circ & 0.0958 \angle 14.97^\circ & 0.7474 \angle 15.74^\circ  \\
0.6958\angle \shortminus166.02^\circ & 0.3328 \angle 12.12^\circ & 0.0937 \angle 7.55^\circ \\
0.3121\angle 6.87^\circ & 0.6946 \angle 4.43^\circ & 0.1524 \angle 178.19^\circ \\
\end{bmatrix}
\end{equation}

The NSE of the implemented circuit with respect to the target is 0.0315. A comparison of the two matrices reveals discrepancies in their entries, primarily due to physical effects discussed in Section \ref{Validation}. 
Nevertheless, the result confirms the feasibility of mapping an IIR system using discrete-time techniques onto a programmable photonic platform. It also validates the broader applicability of the proposed framework to signal processing tasks involving causal and recursive dynamics, especially within photonic signal processing systems.

\begin{table}[tbh]
\centering
\caption{Design parameters for the case (\ref{M_iir}).}
\begin{tabular}{|c|c|c|c|c|c|c|c|c|c|c}
\hline
Components & $\phi_i$ & $K$ & Phase adjustment & $\phi_{o_{i}}$ & $\Delta n_1$ & $\Delta n_2$ & $\Delta n_3$ & $\Delta n_{\text{wg}_{con_i}} $ & $\Delta n_{\text{wg}_{out_i}} $\\
\hline
{DRT$_1$} & -$90^\circ$ & 0.17 & -$54.8^\circ$ & - & 0.0427 & 0.0112 & 0.0259 & - & -\\
\hline
{DRT$_2$} & $90^\circ$ & 0.96 & -$114.2^\circ$ & - & -0.0428 & 0.0016 & 0.0543 & - & -\\
\hline
{DRT$_3$} & -$90^\circ$ & 0.99 & -$115.8^\circ$ & - & 0.0427 & 0.0012 & 0.0551 & - & -\\
\hline
$\text{wg}_{\text{con}_1}$ & - & - & -$11.2^\circ$ & - & - & - & - & 0.0050 & -\\
\hline
$\text{wg}_{\text{con}_2}$ & - & - & -$11.2^\circ$ & - & - & - & - & 0.0050 & -\\
\hline
$\text{wg}_{\text{con}_3}$ & - & - & -$11.2^\circ$ & - & - & - & - & 0.0050 & -\\
\hline
$\text{wg}_{\text{out}_1}$ & - & - & - & -$180^\circ$ & - & - & - & - & 0.0856 \\
\hline
\end{tabular}
\label{table_iir_filter}
\end{table}

\hfill \newline

\section{Conclusion}
In this work, we introduced a scalable and efficient framework for implementing programmable photonic integrated circuits based on DRT resonators. By analytically deriving the transfer function and demonstrating its functional equivalence to conventional DCs, we established a strong foundation for using this structure as tunable building blocks. Our approach enables precise control over amplitude and phase, validated through comprehensive FEM simulations and hybrid modelling techniques. Scalability analysis showed that the NSE in circuits constructed from DRT building blocks increases with the size of the unitary matrix, at a rate higher than that of circuits based on DCs. While PICs based on DRT offer advantages in tunability, their scalability requires further optimisation for large scale applications. The framework successfully mapped both unitary and non-unitary systems, including a practical application as an IIR filter, confirming its potential for broader use in optical signal processing and neuromorphic computing. The presented methodology paves the way for future developments in high-density and reconfigurable photonic circuits that demand compactness, precision, and adaptability. 

\newpage


\printbibliography[heading=bibintoc,title={References}]

@book{Capmany2020,
   author = {José Capmany and Daniel Pérez},
   publisher = {Oxford University Press},
   title = {Programmable integrated photonics},
   year = {2020},
}

@misc{Keysight2017,
   author = {Keysight},
   city = {USA},
   month = {1},
   title = {Application Note RF Design Software Learning Kit Step-By-Step Examples on Using ADS Software for an Introductory RF/ Microwave Course},
   url = {www.keysight.com},
   year = {2017},
}

@INPROCEEDINGS{Sewell2007,
  author={Sewell, P. and Benson, T.M. and Vukovic, A. and Styan, C.},
  booktitle={2007 9th International Conference on Transparent Optical Networks}, 
  title={Adaptive Simulation of Optical ASICs}, 
  year={2007},
  volume={1},
  number={},
  pages={244-249},
  doi={10.1109/ICTON.2007.4296080}}

@article{Vukovic2010,
author = {Ana Vukovic and Phillip Sewell and Trevor M. Benson},
journal = {J. Opt. Soc. Am. A},
number = {10},
pages = {2156--2168},
publisher = {Optica Publishing Group},
title = {Strategies for global optimization in photonics design},
volume = {27},
month = {Oct},
year = {2010},
url = {https://opg.optica.org/josaa/abstract.cfm?URI=josaa-27-10-2156},
doi = {10.1364/JOSAA.27.002156},
}

@article{Piggott2015,
  author = {Piggott, Alexander Y. and Lu, Jesse and Lagoudakis, Konstantinos G. and Petykiewicz, Jan and Babinec, Thomas M. and Vučković, Jelena},
  title = {Inverse design and demonstration of a compact and broadband on-chip wavelength demultiplexer},
  journal = {Nature Photonics},
  volume = {9},
  pages = {374–377},
  year = {2015},
  publisher = {Nature Publishing Group},
  doi = {10.1038/nphoton.2015.69},
  url = {https://www.nature.com/articles/nphoton.2015.69}
}

@incollection{phang2024,
  title = {The optical reservoir computer: a new approach to a programmable integrated optics system based on an artificial neural network},
  author = {Phang, Sendy and Sewell, Phillip D. and Vukovic, Ana D. and Benson, Trevor M.},
  booktitle = {Integrated Optics Volume 2: Characterization, devices and applications},
  pages = {361--380},
  year = {2024},
  publisher = {The Institution of Engineering and Technology},
  doi = {10.1049/pbcs077g_ch12},
  url = {https://digital-library.theiet.org/doi/10.1049/pbcs077g_ch12}
}

@article{Zhuang2015,
author = {Leimeng Zhuang and Chris G. H. Roeloffzen and Marcel Hoekman and Klaus-J. Boller and Arthur J. Lowery},
journal = {Optica},
number = {10},
pages = {854--859},
publisher = {Optica Publishing Group},
title = {Programmable photonic signal processor chip for radiofrequency applications},
volume = {2},
month = {Oct},
year = {2015},
url = {https://opg.optica.org/optica/abstract.cfm?URI=optica-2-10-854},
doi = {10.1364/OPTICA.2.000854},
}

@article{Hu2017,
author = {Ting Hu and Bowei Dong and Xianshu Luo and Tsung-Yang Liow and Junfeng Song and Chengkuo Lee and Guo-Qiang Lo},
journal = {Photon. Res.},
number = {5},
pages = {417--430},
publisher = {Optica Publishing Group},
title = {Silicon photonic platforms for mid-infrared applications},
volume = {5},
month = {Oct},
year = {2017},
url = {https://opg.optica.org/prj/abstract.cfm?URI=prj-5-5-417},
doi = {10.1364/PRJ.5.000417},
}

@ARTICLE{Poulton2019,
  author={Poulton, Christopher Vincent and Byrd, Matthew J. and Russo, Peter and Timurdogan, Erman and Khandaker, Murshed and Vermeulen, Diedrik and Watts, Michael R.},
  journal={IEEE Journal of Selected Topics in Quantum Electronics}, 
  title={Long-Range LiDAR and Free-Space Data Communication With High-Performance Optical Phased Arrays}, 
  year={2019},
  volume={25},
  number={5},
  pages={1-8},
  doi={10.1109/JSTQE.2019.2908555}}

@article{zhang2020,
  title = {Photonic integrated field-programmable disk array signal processor},
  author = {Zhang, Weifeng and Yao, Jianping},
  journal = {Nature Communications},
  volume = {11},
  number = {1},
  pages = {1--9},
  year = {2020},
  publisher = {Nature Publishing Group},
  doi = {10.1038/s41467-019-14249-0},
  url = {https://www.nature.com/articles/s41467-019-14249-0}
}

@article{Carolan2015,
author = {Jacques Carolan  and Christopher Harrold  and Chris Sparrow  and Enrique Martín-López  and Nicholas J. Russell  and Joshua W. Silverstone  and Peter J. Shadbolt  and Nobuyuki Matsuda  and Manabu Oguma  and Mikitaka Itoh  and Graham D. Marshall  and Mark G. Thompson  and Jonathan C. F. Matthews  and Toshikazu Hashimoto  and Jeremy L. O’Brien  and Anthony Laing },
title = {Universal linear optics},
journal = {Science},
volume = {349},
number = {6249},
pages = {711-716},
year = {2015},
doi = {10.1126/science.aab3642},
URL = {https://www.science.org/doi/abs/10.1126/science.aab3642},
eprint = {https://www.science.org/doi/pdf/10.1126/science.aab3642}}

@ARTICLE{Najjar_Amiri2024,
  title    = "Deep photonic network platform enabling arbitrary and broadband
              optical functionality",
  author   = "Najjar Amiri, Ali and Vit, Aycan Deniz and Gorgulu, Kazim and
              Magden, Emir Salih",
  journal  = "Nature Communications",
  volume   =  15,
  number   =  1,
  pages    = "1432",
  month    =  feb,
  year     =  2024
}

@article{Lu2013,
author = {Jesse Lu and Jelena Vu\v{c}kovi\'{c}},
journal = {Opt. Express},
number = {11},
pages = {13351--13367},
publisher = {Optica Publishing Group},
title = {Nanophotonic computational design},
volume = {21},
month = {Jun},
year = {2013},
url = {https://opg.optica.org/oe/abstract.cfm?URI=oe-21-11-13351},
doi = {10.1364/OE.21.013351},
}

@article{Zhang2022,
author = {Guowu Zhang and Dan-Xia Xu and Yuri Grinberg and Odile Liboiron-Ladouceur},
journal = {Photon. Res.},
number = {7},
pages = {1787--1802},
publisher = {Optica Publishing Group},
title = {Experimental demonstration of robust nanophotonic devices optimized by topological inverse design with energy constraint},
volume = {10},
month = {Jul},
year = {2022},
url = {https://opg.optica.org/prj/abstract.cfm?URI=prj-10-7-1787},
doi = {10.1364/PRJ.457066},
}

@ARTICLE{Jia2018,
  title     = "{Inverse-Design} and Demonstration of Ultracompact Silicon
               {Meta-Structure} Mode Exchange Device",
  author    = "Jia, Hao and Zhou, Ting and Fu, Xin and Ding, Jianfeng and Yang,
               Lin",
  journal   = "ACS Photonics",
  publisher = "American Chemical Society",
  volume    =  5,
  number    =  5,
  pages     = "1833--1838",
  month     =  may,
  year      =  2018
}

@ARTICLE{Bogaerts2020,
  title    = "Programmable photonic circuits",
  author   = "Bogaerts, Wim and P{\'e}rez, Daniel and Capmany, Jos{\'e} and
              Miller, David A B and Poon, Joyce and Englund, Dirk and
              Morichetti, Francesco and Melloni, Andrea",
  journal  = "Nature",
  volume   =  586,
  number   =  7828,
  pages    = "207--216",
  month    =  oct,
  year     =  2020
}

@ARTICLE{Xu2022,
  title    = "Self-calibrating programmable photonic integrated circuits",
  author   = "Xu, Xingyuan and Ren, Guanghui and Feleppa, Tim and Liu, Xumeng
              and Boes, Andreas and Mitchell, Arnan and Lowery, Arthur J",
  journal  = "Nature Photonics",
  volume   =  16,
  number   =  8,
  pages    = "595--602",
  month    =  aug,
  year     =  2022
}

@ARTICLE{Perez2020,
  title    = "Multipurpose self-configuration of programmable photonic circuits",
  author   = "P{\'e}rez-L{\'o}pez, Daniel and L{\'o}pez, Aitor and
              DasMahapatra, Prometheus and Capmany, Jos{\'e}",
  journal  = "Nature Communications",
  volume   =  11,
  number   =  1,
  pages    = "6359",
  month    =  dec,
  year     =  2020
}

@ARTICLE{Perez2017,
  title    = "Multipurpose silicon photonics signal processor core",
  author   = "P{\'e}rez, Daniel and Gasulla, Ivana and Crudgington, Lee and
              Thomson, David J and Khokhar, Ali Z and Li, Ke and Cao, Wei and
              Mashanovich, Goran Z and Capmany, Jos{\'e}",
  journal  = "Nature Communications",
  volume   =  8,
  number   =  1,
  pages    = "636",
  month    =  sep,
  year     =  2017
}

@article{Yi2021,
    author = {Yi, Dan and Wang, Yi and Tsang, Hon Ki},
    title = {Multi-functional photonic processors using coherent network of micro-ring resonators},
    journal = {APL Photonics},
    volume = {6},
    number = {10},
    pages = {100801},
    year = {2021},
    month = {10},
    issn = {2378-0967},
    doi = {10.1063/5.0062865},
    url = {https://doi.org/10.1063/5.0062865},
    eprint = {https://pubs.aip.org/aip/app/article-pdf/doi/10.1063/5.0062865/16694225/100801\_1\_online.pdf},
}

@article{Talib2025,
  author    = {Hussein Talib and Phillip D. Sewell and Ana Vukovic and Sendy Phang},
  title     = {Photonic circuit of arbitrary non-unitary systems},
  journal   = {Optical and Quantum Electronics},
  volume    = {57},
  pages     = {99},
  year      = {2025},
  doi       = {10.1007/s11082-024-07957-5},
  url       = {https://doi.org/10.1007/s11082-024-07957-5}
}

@article{Clements2016,
author = {William R. Clements and Peter C. Humphreys and Benjamin J. Metcalf and W. Steven Kolthammer and Ian A. Walmsley},
journal = {Optica},
number = {12},
pages = {1460--1465},
publisher = {Optica Publishing Group},
title = {Optimal design for universal multiport interferometers},
volume = {3},
month = {Dec},
year = {2016},
url = {https://opg.optica.org/optica/abstract.cfm?URI=optica-3-12-1460},
doi = {10.1364/OPTICA.3.001460},
}

@article{reck_1994,
  title = {Experimental realisation of any discrete unitary operator},
  author = {Reck, Michael and Zeilinger, Anton and Bernstein, Herbert J. and Bertani, Philip},
  journal = {Phys. Rev. Lett.},
  volume = {73},
  issue = {1},
  pages = {58--61},
  numpages = {0},
  year = {1994},
  month = {Jul},
  publisher = {American Physical Society},
  doi = {10.1103/PhysRevLett.73.58},
  url = {https://link.aps.org/doi/10.1103/PhysRevLett.73.58}
}

@article{Miller:13,
author = {David A. B. Miller},
journal = {Photon. Res.},
number = {1},
pages = {1--15},
publisher = {Optica Publishing Group},
title = {Self-configuring universal linear optical component},
volume = {1},
month = {Jun},
year = {2013},
url = {https://opg.optica.org/prj/abstract.cfm?URI=prj-1-1-1},
doi = {10.1364/PRJ.1.000001},
}

@article{Fldzhyan:20,
author = {S. A. Fldzhyan and M. Yu. Saygin and S. P. Kulik},
journal = {Opt. Lett.},
number = {9},
pages = {2632--2635},
publisher = {Optica Publishing Group},
title = {Optimal design of error-tolerant reprogrammable multiport interferometers},
volume = {45},
month = {May},
year = {2020},
url = {https://opg.optica.org/ol/abstract.cfm?URI=ol-45-9-2632},
doi = {10.1364/OL.385433},
}

@article{Marchesin:25,
author = {Federico Marchesin and Matĕj Hejda and Tzamn Melendez Carmona and Stefano Di Carlo and Alessandro Savino and Fabio Pavanello and Thomas Van Vaerenbergh and Peter Bienstman},
journal = {Opt. Express},
number = {2},
pages = {2227--2246},
publisher = {Optica Publishing Group},
title = {Braided interferometer mesh for robust photonic matrix-vector multiplications with non-ideal components},
volume = {33},
month = {Jan},
year = {2025},
url = {https://opg.optica.org/oe/abstract.cfm?URI=oe-33-2-2227},
doi = {10.1364/OE.547206},
}

@article{Diamond:20,
author = {Farhad Shokraneh and Simon Geoffroy-gagnon and Odile Liboiron-Ladouceur},
journal = {Opt. Express},
number = {16},
pages = {23495--23508},
publisher = {Optica Publishing Group},
title = {The diamond mesh, a phase-error- and loss-tolerant field-programmable MZI-based optical processor for optical neural networks},
volume = {28},
month = {Aug},
year = {2020},
url = {https://opg.optica.org/oe/abstract.cfm?URI=oe-28-16-23495},
doi = {10.1364/OE.395441},
}

@article{Addressing:23,
author = {Kaveh (Hassan) Rahbardar Mojaver and Bokun Zhao and Edward Leung and S. Mohammad Reza Safaee and Odile Liboiron-Ladouceur},
journal = {Opt. Express},
number = {15},
pages = {23851--23866},
publisher = {Optica Publishing Group},
title = {Addressing the programming challenges of practical interferometric mesh based optical processors},
volume = {31},
month = {Jul},
year = {2023},
url = {https://opg.optica.org/oe/abstract.cfm?URI=oe-31-15-23851},
doi = {10.1364/OE.489493},
}

@article{mosses2023design,
  title={Design and analysis of on-chip reconfigurable photonic components for photonic multiply and accumulate operation},
  author={Mosses, A. and Prathap, P. M. Joe},
  journal={Optical and Quantum Electronics},
  volume={55},
  number={10},
  pages={934},
  year={2023},
  publisher={Springer},
  doi={10.1007/s11082-023-05200-1},
  url={https://link.springer.com/article/10.1007/s11082-023-05200-1}
}

@article{multifunctional,
url = {https://doi.org/10.1515/nanoph-2018-0051},
title = {Programmable multifunctional integrated nanophotonics},
title = {},
author = {Daniel Pérez and Ivana Gasulla and José Capmany},
pages = {1351--1371},
volume = {7},
number = {8},
journal = {Nanophotonics},
doi = {doi:10.1515/nanoph-2018-0051},
year = {2018},
lastchecked = {2025-04-13}
}

@article{dual-drive,
author = {Daniel P\'{e}rez-L\'{o}pez and Ana M. Gutierrez and Erica S\'{a}nchez and Prometheus DasMahapatra and Jos\'{e} Capmany},
journal = {Opt. Express},
number = {26},
pages = {38071--38086},
publisher = {Optica Publishing Group},
title = {Integrated photonic tunable basic units using dual-drive directional couplers},
volume = {27},
month = {Dec},
year = {2019},
url = {https://opg.optica.org/oe/abstract.cfm?URI=oe-27-26-38071},
doi = {10.1364/OE.27.038071},
}

@book{rabus2020integrated,
  title={Integrated Ring Resonators: A Compendium},
  author={Rabus, Dominik Gerhard and Sada, Cinzia},
  volume={127},
  year={2020},
  publisher={Springer Nature}
}

@article{Anufriev:22,
author = {Gleb Anufriev and David Furniss and Mark Farries and Sendy Phang},
journal = {Opt. Mater. Express},
number = {5},
pages = {1767--1783},
publisher = {Optica Publishing Group},
title = {Non-spectroscopic sensing enabled by an electro-optical reservoir computer},
volume = {12},
month = {May},
year = {2022},
url = {https://opg.optica.org/ome/abstract.cfm?URI=ome-12-5-1767},
doi = {10.1364/OME.449036},
}

@manual{comsol2023,
  title        = {{COMSOL Multiphysics\textsuperscript{\textregistered} v6.2, Wave Optics Module User's Guide}},
  organization = {COMSOL AB},
  address      = {Stockholm, Sweden},
  year         = {2023},
  note         = {\url{https://www.comsol.com}},
}

@misc{talib2025owtnm,
  author       = {Talib, Hussein},
  title        = {Photonic matrix multiplication using double racetrack resonators lattice},
  howpublished = {Presented at the 31st International Workshop on Optical Wave \& Waveguide Theory and Numerical Modelling (OWTNM)},
  month        = April,
  year         = 2025,
  note         = {Nottingham Trent University, Nottingham, UK},
  url          = {https://iop.eventsair.com/owtnm2025/}
}

@article{valdez2025high,
  title={High-contrast nulling in photonic meshes through architectural redundancy},
  author={Valdez, Carson G and Sun, Zhanghao and Kroo, Anne R and Miller, David AB and Solgaard, Olav},
  journal={Optics Letters},
  volume={50},
  number={11},
  pages={3660--3663},
  year={2025},
  publisher={Optica Publishing Group}
}

@article{fandino2017monolithic,
  title={A monolithic integrated photonic microwave filter},
  author={Fandi{\~n}o, Javier S and Mu{\~n}oz, Pascual and Dom{\'e}nech, David and Capmany, Jos{\'e}},
  journal={Nature Photonics},
  volume={11},
  number={2},
  pages={124--129},
  year={2017},
  publisher={Nature Publishing Group UK London}
}




\end{document}